\documentclass[12pt]{article}

\ifx\pdfoutput\undefined
\usepackage[dvips,bookmarks]{hyperref}
\else
\usepackage{hyperref}
\fi
\hypersetup{colorlinks=false,bookmarksopen,bookmarksnumbered,citecolor=blue,
   pdfstartview=FitH}

\usepackage[dvips]{graphicx}
\usepackage{latexsym}
\usepackage{amssymb,amsfonts,amsmath}
\usepackage{graphicx} 
\usepackage{indentfirst}

 \usepackage{bbm}

\parskip=\medskipamount

\newcommand {\cF}{{\cal F}}

\newcommand {\cH}{{\cal H}}

\newcommand {\cL}{{\cal L}}
\newcommand {\cM}{{\cal M}}
\newcommand {\cN}{{\cal N}}

\newcommand {\cR}{{\cal R}}

\def\a{\alpha}
\def \bi{\bibitem}

\def\d{\delta}

\def\f{\phi}

\def\G{\Gamma}

\def\q{\theta}

\def\s{\sigma}

\def\z{\zeta}

\def\F{\Phi}
\def\J{\Psi}

\def\S{\Sigma}
\def\U{\Upsilon}

\newcommand{\ve}{\varepsilon}                            

\newcommand{\pa}{\partial}                           
\newcommand{\hf}{\frac12}

\newcommand{\be}{\begin{equation}}
\newcommand{\ee}{\end{equation}}
\newcommand{\bea}{\begin{eqnarray}}
\newcommand{\eea}{\end{eqnarray}}
\newcommand{\non}{\nonumber}

\def\dt#1{{\buildrel {\hbox{\LARGE .}} \over {#1}}}    

\newcommand{\bm}[1]{\mbox{\boldmath$#1$}}

\def\double #1{#1{\hbox{\kern-2pt $#1$}}}

\begin{document}

\begin{titlepage}

\begin{flushright}
November, 2008\\
\end{flushright}
\vspace{5mm}

\begin{center}
{\Large \bf  Chiral formulation for  
hyperk\"ahler sigma-models on}\\
{\Large \bf  cotangent bundles of symmetric  spaces  }
\end{center}

\begin{center}

{\large  
Sergei M. Kuzenko\footnote{{kuzenko@cyllene.uwa.edu.au}}
and Joseph Novak\footnote{novakj02@tartarus.uwa.edu.au}} \\
\vspace{5mm}

\footnotesize{
{\it School of Physics M013, The University of Western Australia\\
35 Stirling Highway, Crawley W.A. 6009, Australia}}  
~\\

\vspace{2mm}

\end{center}
\vspace{5mm}

\begin{abstract}
\baselineskip=14pt
Starting with the projective-superspace off-shell formulation for four-dimensional 
$\cN=2$ supersymmetric sigma-models on cotangent bundles of {\it arbitrary} 
Hermitian symmetric spaces, 
their on-shell description in terms of $\cN=1$ chiral superfields is developed.  
In particular, we derive a universal representation for the hyperk\"ahler potential 
in terms of the curvature of the symmetric base space. 
Within the tangent-bundle formulation for such sigma-models, 
completed recently in arXiv:0709.2633 and realized in terms of $\cN=1$ chiral 
and complex linear superfields, 
we give a new universal formula for the superspace Lagrangian.
A closed form expression is also derived for the K\"ahler potential 
of an arbitrary Hermitian symmetric space in K\"ahler normal coordinates.
\end{abstract}
\vspace{1cm}

\vfill
\end{titlepage}

\newpage
\renewcommand{\thefootnote}{\arabic{footnote}}
\setcounter{footnote}{0}

\section{Introduction}
\setcounter{equation}{0}
Ten years ago, it was noticed \cite{K-double,GK1}, 
using the projective-superspace techniques \cite{LR}, that 
the general four-dimensional $\cN=1$ supersymmetric 
nonlinear sigma-model \cite{Zumino}
\be
S[\F, \bar \F] =  \int 
{\rm d}^4 x \,{\rm d}^4\q
\, K(\Phi^{I},
 {\bar \Phi}{}^{\bar{J}})  ~,
\label{nact4}
\ee
with $K$ the  K\"ahler potential of a K\"ahler manifold $\cM$, 
admits an off-shell $\cN=2$ extension formulated in $\cN=1$
superspace as follows:
\bea
S[\U, \breve{\U}]  =  
\frac{1}{2\pi {\rm i}} \, \oint \frac{{\rm d}\z}{\z} \,  
 \int  {\rm d}^4 x \,{\rm d}^4\q
  \, K \big( \U^I (\z), \breve{\U}^{\bar{J}} (\z)  \big) ~.
\label{nact} 
\eea
Here $\z \in {\mathbb C} \setminus {0}$ is an auxiliary complex variable, 
and the dynamical variables $\U(\z)$ and $\breve{\U} (\z) $ comprise 
an infinite set of ordinary $\cN=1$ superfields:
\be
 \U (\z) = \sum_{n=0}^{\infty}  \, \U_n \z^n = 
\F + \S \,\z+ O(\z^2) ~,\qquad
\breve{\U} (\z) = \sum_{n=0}^{\infty}  \, {\bar
\U}_n
 (-\z)^{-n}~,
\label{exp}
\ee
with $\F$  chiral, $\S$  complex linear, 
\be
{\bar D}_{\dt{\a}} \F =0~, \qquad \qquad {\bar D}^2 \S = 0 ~,
\label{chiral+linear}
\ee
and the remaining component, $\U_2, \U_3, \dots$,  being unconstrained complex 
superfields.\footnote{In the terminology of \cite{G-RLRWvU}, 
the superfields $\U(\z)$ and $\breve{\U} (\z) $ realize a polar hypermultiplet.
The most general $\cN=2$ supersymmetric sigma-model couplings of polar hypermultiplets 
\cite{LR} are obtained from (\ref{nact}) by allowing $K$ to depend explicitly on $\z$, 
$K \big( \U, \breve{\U}) \to K \big( \U, \breve{\U}, \z \big)$.  
A geometric interpretation of such generalized couplings has recently been discussed in
\cite{LR2}.}
The latter enter the action without derivatives, and therefore they form 
an infinite set of auxiliary superfields.
As pointed out in \cite{K-double}, 
the $\cN=2$ supersymetric sigma-model  (\ref{nact}) 
inherits  all the geometric features of
its $\cN=1$ predecessor (\ref{nact4}), that is  
properly realised K\"ahler symmetry and 
invariance under holomorphic reparametrizations of the K\"ahler manifold. 
The latter property implies that the variables $(\F^I, \S^J)$ parametrize the tangent 
bundle $T\cM$ of the K\"ahler manifold $\cM$ \cite{K-double}. 

The auxiliary superfields $\U_2, \U_3, \dots$ can in principle be integrated out, 
at least in perturbation theory, and then the action (\ref{nact}) turns into  \cite{GK1}
\bea
S_{{\rm tb}}[\F,  \S]  
&=& \int {\rm d}^4 x \,{\rm d}^4\q\, \Big\{\,
K \big( \F, \bar{\F} \big)+   \cL\big(\F, \bar \F, \S , \bar \S \big) \Big\}~, 
\label{act-tab}
\eea
where the second term in the Lagrangan looks like 
\bea
 \cL &=&
\sum_{n=1}^{\infty} \cL^{(n)} ~, \qquad
\cL^{(n)} =
\cL_{I_1 \cdots I_n {\bar J}_1 \cdots {\bar 
J}_n }  \big( \F, \bar{\F} \big) \S^{I_1} \dots \S^{I_n} 
{\bar \S}^{ {\bar J}_1 } \dots {\bar \S}^{ {\bar J}_n }~.~~~
\eea
Here $\cL_{I {\bar J} }=  - g_{I \bar{J}} \big( \F, \bar{\F}  \big) $, 
while the tensors $\cL_{I_1 \cdots I_n {\bar J}_1 \cdots {\bar 
J}_n }$ for  $n>1$ 
are functions of the Riemann curvature $R_{I {\bar 
J} K {\bar L}} \big( \F, \bar{\F} \big) $ and its covariant 
derivatives.

The theory with action (\ref{act-tab}) possesses a dual formulation. 
It can be obtained by  considering the first-order action
\bea
S=  \int 
{\rm d}^4 x \,{\rm d}^4\q\, \Big\{\,
K \big( \F, \bar{\F} \big)+  \cL
\big(\F, \bar \F, \S , \bar \S \big)
+\J_I \,\S^I + {\bar \J}_{\bar I} {\bar \S}^{\bar I} 
\Big\}~,
\label{f-o}
\eea
where the tangent vector $\S^I$ is now  complex unconstrained, 
while the one-form 
$\Psi_I$ is chiral, ${\bar D}_{\dt \a} \J_I =0$.
Integrating out $\S$'s and their conjugates gives
\bea
S_{{\rm ctb}}[\F,  \J]  
&=& \int 
{\rm d}^4 x \,{\rm d}^4\q
\, \Big\{\,
K \big( \F, \bar{\F} \big)+    
\cH \big(\F, \bar \F, \J , \bar \J \big)\Big\}~,
\label{act-ctb}
\eea
where the second term in the Lagrangian is 
\bea
\cH
&=& \sum_{n=1}^{\infty} 
\cH^{(n)}~, \quad  
\cH^{(n)} 
=
\cH^{I_1 \cdots I_n {\bar J}_1 \cdots {\bar 
J}_n }  \big( \F, \bar{\F} \big) \J_{I_1} \dots \J_{I_n} 
{\bar \J}_{ {\bar J}_1 } \dots {\bar \J}_{ {\bar J}_n } ~,
\label{h}
\eea
with $\cH^{I {\bar J}} =g^{I {\bar J}} \big( \F, \bar{\F} \big) $.
The variables $(\F^I, \J_J)$ parametrize the cotangent 
bundle $T^* \cM$ of the K\"ahler manifold $\cM$ \cite{GK1}.
Since the theory with action (\ref{act-ctb}) is $\cN=2$ supersymmetric
and realized in terms of chiral superfields, 
the Lagrangian in (\ref{act-ctb}) constitutes the hyperk\"ahler potential 
for (in general, an open domain of the zero section of) $T^* \cM$, 
in accordance with \cite{A-GF}.
If $\cM$ is a compact Hermitian symmetric space, then the hyperk\"ahler 
structure turns out to be  globally defined on $T^* \cM$.

The problem of explicit computation
of $\cL \big(\F, \bar \F, \S , \bar \S \big)$ and
$\cH \big(\F, \bar \F, \J , \bar \J \big)$ from the off-shell sigma-model  (\ref{nact}) 
was addressed in a series of papers \cite{GK1,GK2,AN,AKL1,AKL2}
for the case when $\cM$ is  a Hermitian symmetric space. 
The method\footnote{The method was introduced in \cite{GK1} and 
illustrated on the example of $\cM ={\mathbb C}P^1$. 
The case of ${\mathbb C}P^n$ was worked out in \cite{GK2,AN}.
The  classical compact symmetric spaces $U(n+m)/U(n) \times U(m)$, $SO(2n)/U(n)$, 
$SP(n)/U(n)$ and $SO(n+2)/SO(n) \times SO(2)$, as well as their non-compact 
versions, were worked out in \cite{AKL1}. The tangent-bundle formulation for 
$SO(n+2)/SO(n) \times SO(2)$ was given for the first time in \cite{AN}.} 
used in  \cite{GK1,GK2,AN,AKL1}
essentially exploited the property of such a manifold $\cM$ to be a homogeneous 
space with respect to an appropriate Lie group of holomorphic isometries.
Being perfectly viable, such a setting has a minor disadvantage in the sense that it requires 
a separate consideration for different Hermitian symmetric spaces, on case by case basis.
In particular, this method becomes somewhat cumbersome in 
the case of exceptional symmetric spaces including the compact ones 
$E_6/ SO(10) \times U(1)$  and $E_7/ E_6 \times U(1)$. 
To address the latter spaces, the conceptual set-up was changed in Ref. \cite{AKL2},
which  built on the property of any Hermitian symmetric spaces 
that its curvature tensor is covariantly constant,
\be
\nabla_L  R_{I_1 {\bar  J}_1 I_2 {\bar J}_2}
= {\bar \nabla}_{\bar L} R_{I_1 {\bar  J}_1 I_2 {\bar J}_2} =0~.
\label{covar-const}
\ee
In conjunction with supersymmetry considerations, 
this idea allowed the authors  of \cite{AKL2}
to derive the following closed form expression for the tangent-bundle 
Lagrangian:
\bea
\cL \big(\F, \bar \F, \S , \bar \S \big)
&=& - g_{I \bar{J}} 
 {\bar \S}^{\bar{J}} \,
\frac{ {\rm e}^{\cR_{\S,{\bar \S}}} -1}{ \cR_{\S,{\bar \S} }}\,
 \S^I   ~, \qquad 
 \cR_{\S,{\bar \S}} = -\hf \S^K {\bar \S}^{ {\bar L} } \,
R_{K {\bar L} I }{}^J   \,\S^I \frac{\pa}{\pa \S^J}~.~~~~
\label{closed}
\eea
Using this representation, the case of $E_6/ SO(10) \times U(1)$ was worked out 
in \cite{AKL2} for the first time.\footnote{The tangent-bundle formulation for 
$E_7/ E_6 \times U(1)$ was sketched in \cite{K-hyper2}.}
However, no universal closed form 
expression for $\cH \big(\F, \bar \F, \J , \bar \J \big)$ 
was found in  \cite{AKL2}. One of the aims of the present work is to fill this gap.

This paper is organized as follows. In section 2, we derive an alternative closed form 
expression for $\cL \big(\F, \bar \F, \S , \bar \S \big)$ which differs from (\ref{closed}).
The specific feature of this new representation is that the curvature tensor appears in it as 
a matrix, unlike the differential operator in eq. (\ref{closed}).
In section 3, we derive  $\cH \big(\F, \bar \F, \J , \bar \J \big)$ in a closed form.
Finally, the appendix is devoted to deriving a  closed form expression for the K\"ahler potential 
of an arbitrary Hermitian symmetric space in so-called K\"ahler normal coordinates
(or Bochner's canonical coordinates)
\cite{Bochner,Calabi}.
In the main body of the paper, the K\"ahler manifold $\cM$ is only assumed to obey 
eq. (\ref{covar-const}).

A few words are in order regarding the content of the appendix. 
Recently, an intimate connection was pointed out in  Ref. \cite{K-hyper2}
between the tangent-bundle Lagrangian $\cL \big(\F, \bar \F, \S , \bar \S \big)$
in (\ref{act-tab})  and the K\"ahler potential $K(\f,\bar \f)$ for $\cM$ given 
in K\"ahler normal coordinates $\f$ with origin at $\F$. 
In the symmetric case, eq. (\ref{covar-const}), this correspondence 
is as follows:
\be
\cL \big( \S , \bar \S \big) = K(\f \to -\S\,,\,\bar \f \to \bar \S)~.
\label{correspondence}
\ee
The derivation of eq. (\ref{closed}) in \cite{AKL2}, or the equivalent 
representation (\ref{closed2}) below, are based on supersymmetry
consideration.  Due to (\ref{correspondence}), there should exist a purely 
geometric way of deriving analogues of the representations (\ref{closed})
and  (\ref{closed2})
for $K(\f,\bar \f)$. It is presented in the appendix.

\section{Tangent-bundle formulation}
\setcounter{equation}{0}

The Lagrangian $\cL \big(\F, \bar \F , \S , \bar \S \big)$ obeys the 
first-order linear differential equation \cite{AKL2}
\bea
\hf \S^K \S^L\,R_{K {\bar J} L }{}^I\, \cL_I 
+  \cL_{\bar J}  +g_{I \bar{J}}\, \S^I =0~,
\qquad \cL_I :=  \frac{\pa \cL}{\pa \S^I}~ 
\label{fd2}
\eea
and its conjugate. As demonstrated in \cite{AKL2}, 
this equation expresses the fact that the theory
 (\ref{act-tab}) is $\cN=2$ supersymmetric. It can be shown 
 that this equation is identically satisfied by the function (\ref{closed}).
 A different representation for this  solution is provided below.

It proves robust to rewrite (\ref{fd2}) in a matrix form. For this purpose, we
introduce the following matrices:
\bea
{\bm R}_{\S,\bar \S}
:=\left(
\begin{array}{cc}
0 & (R_\S)^I{}_{\bar J}\\
(R_{\bar \S})^{\bar I}{}_J &0 
\end{array}
\right)~, 
\qquad (R_\S)^I{}_{\bar J}:=\hf R_K{}^I{}_{L \bar J}\, \S^K \S^L~, 
\quad (R_{\bar \S})^{\bar I}{}_J := \overline{(R_\S)^I{}_{\bar J}}~~~~~
\label{R-Sigma}
\eea
and
\bea
{\bm g}
:=\left(
\begin{array}{cc}
0 & g_{I \bar J}\\
g_{{\bar I}J} &0 
\end{array}
\right)
\equiv \left(
\begin{array}{cc}
0 & \hat{g}\\
\check{g} &0 
\end{array}
\right)~. 
\eea
Then  eq. (\ref{fd2}) is equivalent to  
\bea
\left(
\begin{array}{c}
\cL_I\\
\cL_{\bar I} 
\end{array}
\right) = - {\bm g} \Big( {\mathbbm 1} + {\bm R}_{\S,\bar \S}\Big)^{-1} 
\left(
\begin{array}{c}
\S^I\\
\S^{\bar I} 
\end{array}
\right) ~.
\label{der1}
\eea
This relation actually  allows one to determine $\cL$ by taking into account the identities
\bea
\S^I \cL_I =  {\bar \S}^{\bar I}  \cL_{\bar I} 
=   \sum_{n=1}^{\infty}  n \,\cL^{(n)}~. 
\eea
One then obtains 
\bea
\cL \big( \F, \bar \F,  \S,  {\bar \S} \big) &=& - \hf {\bm \S}^{\rm T} {\bm g} \,
\frac{ \ln \big( {\mathbbm 1} + {\bm R}_{\S,\bar \S}\big)}{\bm R_{\S,\bar \S}}
\, {\bm \S}~, \qquad 
{\bm \S} :=\left(
\begin{array}{c}
\S^I\\
{\bar \S}^{\bar I} 
\end{array}
\right) ~.
\label{closed2}
\eea
It also follows from (\ref{der1}) that the following composites
\bea
F^{(2k+2)}&:= &\S^{\rm T} \hat{g} (R_{\bar \S} R_\S)^k R_{\bar \S} \S
= {\S}^\dagger \check{g} (R_\S R_{\bar \S} )^k R_{\S} {\bar \S}~, \qquad k=0,1, 2,\dots
\non \\
F^{(2k+1)}&:= &\S^{\rm T} \hat{g} (R_{\bar \S} R_\S)^k  {\bar \S}
= { \S}^\dagger \check{g} (R_\S R_{\bar \S} )^k \S~, ~~\qquad \qquad k=0,1, 2,\dots
\eea
which appear in the Taylor expansion of $\cL \big( \F, \bar \F,  \S,  {\bar \S} \big)$, 
have the properties
\bea
F^{(2k+2)}_I &=& (2k+2) \hat{g} (R_{\bar \S} R_\S)^k R_{\bar \S} \S~, 
\qquad 
F^{(2k+1)}_I= (2k+1)  \hat{g} (R_{\bar \S} R_\S)^k  {\bar \S}~.
\label{deri1}
\eea
Eq. (\ref{closed2}) constitutes our new closed form expression for 
$\cL  $, compare with (\ref{closed}).

In \cite{AKL2}, it was conjectured that $\cL \big( \F, \bar \F,  \S,  {\bar \S} \big) $
can be represented in the form 
\bea
\cL \big( \F, \bar \F,  \S,  {\bar \S} \big) &=& -{ \S}^\dagger \hat{ g} \,
\frac{ \ln \big( {\mathbbm 1} + {\mathbb R}_{\S,\bar \S}\big)}{ {\mathbb R}_{\S,\bar \S}}
\, { \S}~, \qquad 
({\mathbb R}_{\S, \bar \S})^{ I}{}_J := \hf R_J{}^I{}_{K\bar L} \S^K {\bar \S}^{\bar L}~
\label{closed3}
\eea
which differs from (\ref{closed2}).
The validity of this representation was checked in \cite{AKL2}   for the followings two
choices of $\cM$: 
(i) ${\mathbb C}P^n$; and (ii) $SO(n+2)/SO(n) \times SO(2)$.
Unlike the representation  (\ref{closed2}),
we still do not have a proof that  (\ref{closed3}) holds in general
(however, see comments at the end of the next section).

Using the correspondence (\ref{correspondence})
and K\"ahler normal coordinate considerations (see the appendix), 
one can derive an alternative second-order differential equation enjoyed by $\cL$: 
\be
\cL_{IJ} = \hf R_{I}{}^K{}_J{}^L \, \cL_K \cL_L~. 
\ee

\section{Cotangent-bundle formulation}
\setcounter{equation}{0}

The ``Hamiltonian'' $\cH(\F, \bar \F, \J , \bar \J ) $ obeys 
the nonlinear differential equation \cite{AKL2}
\be
\cH^I \,  g_{I {\bar J}} - \hf \, \cH^K\cH^L \,  R_{K {\bar J} L}{}^I \,\J_I =
{\bar \J}_{ \bar J} ~, 
\qquad \cH^I  = \frac{\pa \cH}{\pa \J_I} ~.
\label{cot-eq}
\ee
This equation immediately follows from (\ref{fd2}) if one makes use of the 
standard properties of the Legendre transformation. 
Alternatively, eq. (\ref{cot-eq}) is equivalent to the condition that the cotangent-bundle 
action (\ref{act-ctb}) is $\cN=2$ supersymmetric \cite{AKL2}.  
The hidden SUSY transformation, which is not manifest in the $\cN=1$  
superspace formulation, is  \cite{AKL2}:
\bea
\d \F^I &=&\hf {\bar D}^2 \big\{ \overline{\ve \q} \, \cH^I  \big(\F, \bar \F, \J , \bar \J \big) \big\} ~, \non \\
\d \J_I &=&- \hf {\bar D}^2 \Big\{ \overline{\ve \q} \, K_I \big( \F, \bar{\F})  \Big\}
+\hf {\bar D}^2 \Big\{ \overline{\ve \q} \, \G^K_{~IJ} \big( \F, \bar{\F} \big)\,
\cH^J  \big(\F, \bar \F, \J , \bar \J \big) \Big\} \,\J_K~,
\label{SUSY4}
\eea
with $\G^K_{~IJ} $ the Christoffel symbols for the K\"ahler metric.
The nonlinearity of (\ref{cot-eq}) makes it  more difficult to solve than  (\ref{fd2}).
Below we provide the solution to eq. (\ref{cot-eq}). 

Equation (\ref{cot-eq}) implies 
\bea
\J_I \cH^I  -  \, \cH^K\cH^L \,  (R_\J )_{K L} =
 g^{I {\bar J}} \J_I {\bar \J}_{\bar J}~, \qquad 
 (R_\J )_{K L}:= \hf R_{K}{}^I{}_{L}{}^J \,\J_I \J_J~. 
 \label{cot-eq2}
\eea
Due to the identities
\bea
\J_I \cH^I =  {\bar \J}_{\bar I}  \cH^{\bar I} 
=   \sum_{n=1}^{\infty}  n \,\cH^{(n)}~, 
\eea
the latter equation 
is equivalent to the following infinite system of equations
\bea
\cH^{(1)}= g^{I {\bar J}} \J_I {\bar \J}_{\bar J}~, \qquad 
n\cH^{(n)} - \sum_{p=1}^{n-1} \cH^{(p)\, K} (R_\J )_{K L}{} \cH^{(n-p)\,L}=0~, 
\quad n\geq 2~.
 \label{cot-eq3}
\eea
It is clear that the contributions $\cH^{(2)}, \cH^{(3)}, \dots$, can be uniquely 
determined, order by order in perturbation theory,  using  the  equations derived.

To solve (\ref{cot-eq3}), it is useful to introduce a matrix associated with the Riemann tensor 
\bea
{\bm R}_{\J,\bar \J}
:=\left(
\begin{array}{cc}
0 & (R_\J)_I{}^{\bar J}\\
(R_{\bar \J})_{\bar I}{}^J &0 
\end{array}
\right)~, 
\qquad (R_\J)_I{}^{\bar J}= (R_\J)_{IK} \,g^{K \bar J}~,
\eea
as well as a family of building blocks
\bea
G{}^{(2k+2)}&:= &\J^{\rm T} \hat{g}^{-1}  (R_{\bar \J} R_\J)^k R_{\bar \J} \J
= { \J}^\dagger \check{g}^{-1} (R_\J R_{\bar \J} )^k R_{\J} {\bar \J}~, \qquad k=0,1, 2,\dots
\non \\
G{}^{(2k+1)} &:= &\J^{\rm T} \hat{g}^{-1} (R_{\bar \J} R_\J)^k  {\bar \J}
= { \J}^\dagger \check{g}^{-1} (R_\J R_{\bar \J} )^k \J~, \qquad \qquad ~~k=0,1, 2,\dots~.
\eea
Their partial derivatives can be read off from (\ref{deri1})
\bea
G{}^{(2k+2)\,I} &:= & (2k+2)\hat{g}^{-1}  (R_{\bar \J} R_\J)^k R_{\bar \J} \J
=(2k+2)\J^{\rm T} \hat{g}^{-1}  (R_{\bar \J} R_\J)^k R_{\bar \J} ~, \non \\
G {}^{(2k+1)\,I}&:= &(2k+1)  \hat{g}^{-1} (R_{\bar \J} R_\J)^k  {\bar \J}
= (2k+1)\J^\dagger \hat{g}^{-1} (R_\J R_{\bar \J} )^k ~. 
\eea 
Now, if one introduces an ansatz 
\bea
\cH^{(n)} =  c_n\,G^{(n)} ~,\qquad n\geq 2
\eea
with $c_n$ numerical coefficients, 
the equations (\ref{cot-eq3}) turn into 
the following system of quadratic algebraic equations: 
\bea
n \,c_{n} - \sum_{p=1}^{n-1}  p (n-p) \,c_{p} c_{n-p}=0~, \qquad c_1=1~.
 \label{cot-eq4}
\eea

The algebraic equations (\ref{cot-eq4}) are universal and independent of the symmetric 
space $\cM$ chosen. Therefore, their solution can be 
deduced by considering any useful choice of $\cM$, 
for which $\cH$ is known, 
say the projective space ${\mathbb C}P^n$ first considered by Calabi
\cite{Calabi2}. 
This observation immediately leads to the solution
\bea
\cH(\F, \bar \F, \J , \bar \J ) = \hf {\bm \J}^{\rm T}{\bm g}^{-1}  \cF \Big( - {\bm R}_{\J,\bar \J} \Big)\, 
{\bm \J} ~, \qquad 
{\bm \J} :=\left(
\begin{array}{c}
\J_I\\
{\bar \J}_{\bar I} 
\end{array}
\right) ~,
\label{hyperkahler-potential}
\eea
where
\be
\cF(x) = \frac{1}{x} \,\Big\{ \sqrt{1+4x} -1 -\ln \frac{1+ \sqrt{1+4x} }{2} \Big\}~, 
\qquad \cF(0)=1~.
\ee
Eq. (\ref{hyperkahler-potential}) is the main result of this work. 

To write down the supersymmetry transformation (\ref{SUSY4}) explicitly, 
we need to compute $\cH^I$ and its conjugate. Direct calculations give
\bea
\left(
\begin{array}{c}
\cH^I\\
\cH^{\bar I} 
\end{array}
\right) = -\hf {\bm g}^{-1} \frac{ 
\sqrt{ {\mathbbm 1} -4 {\bm R}_{\J,\bar \J}}-{\mathbbm 1}}
{{\bm R}_{\J,\bar \J}} \, 
\left(
\begin{array}{c}
\J_I\\
{\bar \J}_{\bar I} 
\end{array}
\right) ~.
\eea

Our derivation of the hyperk\"ahler potential for $T^*\cM$, 
\bea
K(\F, \bar \F) +  \hf {\bm \J}^{\rm T}{\bm g}^{-1}  \cF \Big( - {\bm R}_{\J,\bar \J} \Big)\, 
{\bm \J} ~, 
\eea
was based on the considerations of extended supersymmetry. 
In the mathematical literature, there exists a different representation
for  $\cH(\F, \bar \F, \J , \bar \J )$ \cite{BG}:
\bea
\cH(\F, \bar \F, \J , \bar \J ) =  { \J}^\dagger \check{g}^{-1}  \cF \Big( - {\mathbb R}_{\J,\bar \J} \Big)\, 
{ \J} ~, \qquad 
({\mathbb R}_{\J, \bar \J})_{I}{}^J := \hf R_I{}^{J \bar K L} \J_L {\bar \J}_{\bar K}~.
\eea
This unified formula was derived by Biquard and  Gauduchon
by using purely algebraic means involving 
the root theory for Hermitian symmetric spaces. 
It should be pointed out that the operator ${\mathbb R}_{\J, \bar \J}$ above is 
related to ${\mathbb R}_{\S, \bar \S}$  appearing in (\ref{closed3}). 
It is worth expecting that similar algebraic arguments 
can be used to prove the validity of (\ref{closed3}) for any Hermitian symmetric space. 

The $\cN=2$ supersymmetric model on $T^*\cM$ constructed above can be generalized 
to include a superpotential consistent with $\cN=2$ supersymmetry. 
In accordance with the analysis in \cite{K-superpotential} (see also \cite{BX}), 
the superpotential is 
\be
{\rm e}^{{\rm i}\s} \int {\rm d}^2 \q\, \J_I  \,{X}^I(\F) ~+~{\rm c.c.} ~,
\ee
where ${\rm e}^{{\rm i}\s} $ is a constant phase factor,
 and $ {X}^I(\F) $ a holomorphic Killing vector of the base 
 K\"ahler manifold $\cM$.
 Similar results hold in five space-time dimensions \cite{K-superpotential,BX}.
\\

\noindent
{\bf Acknowledgements:}\\
This work is supported in part
by the Australian Research Council.

\appendix

\section{K\"ahler normal  coordinates} 
\setcounter{equation}{0}

Let us recall the important notion 
of a {\it canonical} coordinate system for a K\"ahler manifold,
that was introduced by Bochner in 1947 \cite{Bochner} 
and later used by Calabi in the 1950s \cite{Calabi}.\footnote{This
coordinate system was  re-discovered by supersymmetry  practitioners
in the 1980s under the name {\it normal gauge} \cite{GGRS,A-GG,HKLR}.}
In a neighborhood of
any point $p$ of the  K\"ahler manifold $\cM$,  
holomorphic reparametrizations  and K\"ahler transformations
can be used to choose a coordinate system, with origin at $p\in \cM$,
in which the K\"ahler potential takes the form:  
\bea
{K} (\f, \bar \f ) &=&{g}_{I \bar{J}}| \,\f^I {\bar \f}^{\bar J}
+ \sum_{m,n \geq 2}^{\infty}  {K}^{(m,n)} (\f, \bar \f)~,
\non \\
{ K}^{(m,n)} (\f, \bar \f) &:=&
\frac{1}{m! n!}\,
{ K}_{I_1 \cdots I_m {\bar J}_1 \cdots {\bar 
J}_n } | \, \f^{I_1} \dots \f^{I_m} 
{\bar \f}^{ {\bar J}_1 } \dots {\bar \f}^{ {\bar J}_n }~.
\label{normal-gauge} 
\eea
In such a coordinate system, there still remains the freedom to perform 
linear holomorphic reparametrizations which can be used 
to set the metric at the origin to  be ${ g}_{I \bar{J}}= \d_{I \bar{J}}$.
The Taylor coefficients in (\ref{normal-gauge}), ${K}_{I_1 \cdots I_m {\bar J}_1 \cdots {\bar J}_n }| $,
turn out to be tensor functions of the K\"ahler metric,
the Riemann curvature $R_{I {\bar J} K {\bar L}}  $ and its covariant 
derivatives,  all of which are   evaluated at the origin.
In the physics literature, Bochner's canonical coordinates are often called
``K\"ahler normal coordinates''  \cite{HIN}.
We follow this terminology. 
K\"ahler normal coordinates are very useful for various considerations, 
 in particular in the context of the so-called Bergman kernel \cite{DK}. 

In the case of symmetric spaces, 
\bea
\nabla_L  R_{I_1 {\bar  J}_1 I_2 {\bar J}_2}
= {\bar \nabla}_{\bar L} R_{I_1 {\bar  J}_1 I_2 {\bar J}_2} =0
\qquad \Longrightarrow \qquad 
{ K}^{(m,n)}=0~, \quad m\neq n~.
\label{normal-gauge2} 
\eea
The condition of covariant constancy can be rewritten as 
\bea
{\bar \nabla}_{\bar L} R_{I_1}{}^{J_1}{}_ {I_2}{}^{J_2} = 
{\bar \pa}_{\bar L} R_{I_1}{}^{J_1}{}_ {I_2}{}^{J_2} = 0~, 
\eea
and therefore $R_{I_1}{}^{J_1}{}_ {I_2}{}^{J_2} $ is $\bar \f$-independent.
Since 
\be
R_{I_1 {\bar J}_1 I_2 {\bar J}_2}
=K_{I_1 I_2{\bar J}_1 {\bar J}_2}
-g_{M \bar N} \G^M_{I_1I_2} {\bar \G}^{\bar N}_{{\bar J}_1{\bar J}_2} 
=K_{I_1 I_2{\bar J}_1 {\bar J}_2}
-g^{\bar M N} K_{I_1I_2 \bar M} K_{N{\bar J}_1{\bar J}_2}~, 
\ee
and  terms in the Taylor series for the expression on the right involve equal numbers
of $\f$ and $\bar \f$, we conclude\footnote{For the Hermitian symmetric space 
$\cM=G/H$, the constant tensor 
$R_{I_1}{}^{J_1}{}_ {I_2}{}^{J_2} $ can be related to the structure constants of $G$.}
\be
R_{I_1}{}^{J_1}{}_ {I_2}{}^{J_2} =R_{I_1}{}^{J_1}{}_ {I_2}{}^{J_2} |={\rm const}~.
\ee
Then we can write
\bea
\G^M_{I_1I_2, \bar J} = R_{I_1}{}^{M}{}_ {I_2 \bar J} 
= R_{I_1}{}^{M}{}_ {I_2}{}^{N} g_{N\bar J}
=R_{I_1}{}^{M}{}_ {I_2}{}^{N} K_{N\bar J}~, 
\eea
and hence 
\bea
\G^M_{I_1I_2} = R_{I_1}{}^{M}{}_ {I_2}{}^{N} \,K_{N}~.
\eea
Contracting both sides of this equation with the metric, $g_{M\bar Q}$, 
one can arrive at the equation
\bea
K_{I_1I_2} = \hf R_{I_1}{}^{M}{}_ {I_2}{}^{N} \,K_M K_{N}~.
\label{nk1}
\eea

Equation (\ref{nk1}) is highly important, since it makes it possible to uniquely restore 
${K} (\f, \bar \f )$ 
provided its functional form, eqs. 
(\ref{normal-gauge}) and (\ref{normal-gauge2}), 
is taken into account. In particular, using eq. (\ref{nk1}) 
allows one to deduce the following alternative  equation:
\bea
g_{I \bar{J}}|\, \f^I 
+\hf \f^K \f^L R_{K {\bar J} L }{}^I|\, K_I 
=K_{\bar J}~.
\label{nk2}
\eea
For the K\"ahler potential, one obtains 
the following closed form expression:
\bea
K \big( \f, \bar \f   \big) &=& - \hf {\bm \f}^{\rm T} {\bm g}| \,
\frac{ \ln \big( {\mathbbm 1} - {\bm R}_{\f,\bar \f}\big)}{\bm R_{\f,\bar \f}}
\, {\bm \f}~, \qquad 
{\bm \f} :=\left(
\begin{array}{c}
\f^I\\
{\bar \f}^{\bar I} 
\end{array}
\right) ~.
\eea
 Here ${\bm R}_{\f,\bar \f}$ is obtained from (\ref{R-Sigma}) by replacing 
 $\S \to \f$ and $R_K{}^I{}_{L \bar J} \to R_K{}^I{}_{L}{}^J \,g_{J \bar J}|$.

 We should emphasize that our derivation above only relied on eq. (\ref{normal-gauge2}).

\end{document}